\documentclass[useAMS,usenatbib]{mn2e}

\usepackage{graphicx}
\usepackage{amssymb}

\def\ri{{\rm i}}
\def\rf{{\rm f}}

\title[Determination of the free--free Gaunt factor]{Accurate determination of the free--free Gaunt factor\\II -- relativistic Gaunt factors}

\author[P.A.M. van Hoof et al.]
{P. A. M. van Hoof$^1$\thanks{p.vanhoof@oma.be},
G. J. Ferland$^2$,
R. J. R. Williams$^3$,
K. Volk$^4$,
M. Chatzikos$^2$,\newauthor
M. Lykins$^2$,
R. L. Porter$^5$\\
$^1$Royal Observatory of Belgium, Ringlaan 3, B-1180 Brussels, Belgium\\
$^2$Department of Physics \& Astronomy, University of Kentucky, Lexington, KY 40506, USA\\
$^3$AWE plc, Aldermaston, Reading, Berkshire RG7 4PR, UK\\
$^4$Space Telescope Science Institute, 3700 San Martin Drive, Baltimore, MD 21218, USA\\
$^5$Department of Physics \& Astronomy and Center for Simulational Physics, University of Georgia, Athens, GA 30602, USA
}

\begin{document}

\onecolumn

\date{Accepted. Received}

\pagerange{\pageref{firstpage}--\pageref{lastpage}} \pubyear{2014}

\maketitle

\label{firstpage}

\begin{abstract}
When modelling an ionised plasma, all spectral synthesis codes need the
thermally averaged free--free Gaunt factor defined over a very wide range of
parameter space in order to produce an accurate prediction for the spectrum.
Until now no data set exists that would meet these needs completely. We have
therefore produced a table of relativistic Gaunt factors over a much wider
range of parameter space than has ever been produced before. We present tables
of the thermally averaged Gaunt factor covering the range ${}^{10}\log\gamma^2
= -6$ to $10$ and ${}^{10}\log u = -16$ to $13$ for all atomic numbers $Z=1$
through $36$. The data were calculated using the relativistic
Bethe-Heitler-Elwert (BHE) approximation and were subsequently merged with
accurate non-relativistic results in those parts of the parameter space where
the BHE approximation is not valid. These data will be incorporated in the
next major release of the spectral synthesis code {\sc cloudy}. We also
produced tables of the frequency integrated Gaunt factor covering the
parameter space ${}^{10}\log\gamma^2 = -6$ to $+10$ for all values of $Z$
between 1 and 36. All the data presented in this paper are available online.
\end{abstract}


\section{Introduction}

In the past many authors discussed the problem of calculating the line and
continuous spectrum of hydrogenic ions. In this paper we will revisit the
problem of calculating the free--free emission and absorption of such an ion.
The problem is normally described by using the free--free Gaunt factor
\citep{Ga30}, which is a multiplicative factor describing the deviation from
classical theory. For brevity we will sometimes refer to the free--free Gaunt
factor simply as the Gaunt factor below.

Any modern spectral synthesis code, such as {\sc cloudy} \citep{Fe13}, needs
accurate values for the Gaunt factor over a wide range of parameter space.
Unfortunately none of the existing data sets fulfils all the necessary
requirements that {\sc cloudy} imposes. We have therefore undertaken to
calculate a new set of Gaunt factors covering a very wide parameter range.
This range is more than enough to avoid any need for extrapolating the data
(even taking certain possible future extensions of the code into account).
This makes the new tables eminently suitable for {\sc cloudy}, but the data
are presented in such a form that they can also be easily used by other codes
that model the free--free absorption or emission process. In \citet[hereafter
  paper~I]{vh14} we described our calculations of non-relativistic Gaunt
factors using exact quantum-mechanical theory. In this paper we will extend
these calculations into the relativistic regime. This is necessary since {\sc
  cloudy} is designed to handle electron temperatures up to 10~GK.
Relativistic effects are important at these temperatures. However, since {\sc
  cloudy} avoids the temperature regime above 10~GK where electron-positron
pair creation would be important, we do not include this effect in our
calculations.

In Sect.~\ref{sec:gff} we will describe the calculation of the relativistic
thermally averaged Gaunt factors. In Sect.~\ref{totalff} we will calculate
frequency-integrated free--free Gaunt factors and use these to determine the
magnitude of the relativistic effects as a function of temperature. Finally,
in Sect.~\ref{summary} we will present a summary of our results. All the data
presented in this paper are available in electronic form from MNRAS as well as
the {\sc cloudy} web site at http://data.nublado.org/gauntff/.

\section{The free--free Gaunt factor}
\label{sec:gff}

In this paper we will consider the process where an unbound electron is moving
through the Coulomb field of a positively charged nucleus, emitting a photon
of energy $\hbar\omega$ in the process. We will assume that the nucleus is a
point-like charge, which implies that the theory is only strictly valid for
fully stripped ions. It is routinely used as an approximation for other ions
as well though. Unlike paper~I, we will not use exact theory, but will
calculate the Gaunt factors in the Born approximation. The relativistic theory
has been described in \citet{BH34} and will be corrected by a relativistic
version of the \citet{El39} factor. The combined theory is hereafter referred
to as the BHE approximation. This approximation is described in detail in
\citet{It85}, \citet[hereafter N98]{No98}, \citet{It00} and references
therein. Work by \citet{EH69} and \citet{PT75} has shown that the BHE
approximation is an excellent approximation for low values of the atomic
number $Z$. However, this approximation is not valid for low temperatures and
low photon energies because the Coulomb distortion of the wave functions
becomes too large and the Born approximation breaks down. In this regime we
will replace the relativistic data with exact non-relativistic Gaunt factors
(hereafter referred to as NR data). These data were described in Paper~I.
Exactly how we merge the two data sets will be described in more detail in
Sect.~\ref{merging:data}.

\subsection{The Bethe-Heitler-Elwert approximation}

Here we describe the theory needed to calculate the thermally averaged Gaunt
factor in the BHE approximation. We will closely follow the notation shown in
\citet{It85} and N98. We will only repeat the most important definitions
needed for our work.

First we need to define the distribution function for the electron energies.
This is done in its most general form using Fermi-Dirac statistics. This
results in the following normalisation of the distribution function
\begin{equation}
G^-_0(\lambda,\nu) = \lambda^3 \int_{\lambda^{-1}}^\infty \frac{e (e^2 - \lambda^{-2})^{1/2}}{\exp(e-\nu)+1} {\rm d}e.
\label{norm}
\end{equation}
Here $e = E/(kT_{\rm e})$ is a scaled version of the electron energy $E$
(including the rest mass of the electron), $k$ is the Boltzmann constant, and
$T_{\rm e}$ is the electron temperature, which is related to the parameter
$\lambda$ by $\lambda \equiv kT_{\rm e}/(m_{\rm e} c^2)$ where $m_{\rm e}$ is
the electron mass and $c$ the speed of light. We can define a parameter $\eta$
which is a measure for the degeneracy of the electron gas. See Eq.~13 in N98
for a formal definition of this parameter. Throughout this work we will assume
$\eta = -70$ (as was done by N98), which is equivalent to assuming that the
gas is fully non-degenerate and in the low-density limit. This is entirely
appropriate for the conditions that {\sc cloudy} is modelling. See N98 for a
further discussion where they showed that their results were indistinguishable
for $-70 \leq \eta \leq -10$. Using $\eta$, we can define the parameter $\nu$
(which is a scaled version of the electron chemical potential $\mu$ including
the rest mass of the electron) as
\[ \nu = \frac{\mu}{kT_{\rm e}} = \lambda^{-1} + \eta = \lambda^{-1} - 70. \]

Next we need to define the integral of the relativistic cross section weighted
by the electron distribution function. In order to avoid numerical overflow
when multiplying with the factor e$^u$ needed below, we modify the first term
in the integrand given by N98 as follows
\[ {\rm e}^u \, J^-(\lambda,\nu,u,Z) = \int^\infty_{\lambda^{-1}+u} \frac{\pi_\ri^2\,e_\rf}{[\exp(e_\ri-u-\nu)+\exp(-u)]
 \, e_\ri} \left( 1 - \frac{1}{\exp(e_\ri-u-\nu)+1} \right)
\frac{1 - \exp(-2\upi\alpha Ze_\ri\pi_\ri^{-1})}{1 - \exp(-2\upi\alpha Ze_\rf\pi_\rf^{-1})} \]
\[ \phantom{{\rm e}^u \, J^-(\lambda,\nu,u,Z) =} \times
 \left[ \frac{4}{3} - 2e_\rf e_\ri\frac{\pi_\rf^2 + \pi_\ri^2}{\pi_\rf^2\pi_\ri^2} +
   \lambda^{-2} \left( \frac{\beta_\rf e_\ri}{\pi_\rf^3} + \frac{\beta_\ri
     e_\rf}{\pi_\ri^3} - \frac{\beta_\rf\beta_\ri}{\pi_\rf\pi_\ri}\right) +
   2\ln\frac{e_\rf e_\ri + \pi_\rf\pi_\ri - \lambda^{-2}} {\lambda^{-1}u} \right. \]
\begin{equation}
\phantom{{\rm e}^u \, J^-(\lambda,\nu,u,Z) =} \left. \times
 \left( \frac{8 e_\rf e_\ri}{3\pi_\rf\pi_\ri} +
 \frac{u^2}{\pi_\rf^3\pi_\ri^3} \{ e_\rf^2e_\ri^2 +\pi_\rf^2\pi_\ri^2 \}
 + \frac{\lambda^{-2}u}{2\pi_\rf\pi_\ri} \left\{ \frac{e_\rf e_\ri + \pi_\ri^2}{\pi_\ri^3}\beta_\ri
 - \frac{e_\rf e_\ri + \pi_\rf^2}{\pi_\rf^3}\beta_\rf + \frac{2\,u\,e_\rf e_\ri}{\pi_\rf^2\pi_\ri^2}
 \right\} \right) \right] {\rm d}e_\ri,
\label{cs:int}
\end{equation}
with
\[ u = \frac{\hbar\omega}{kT_{\rm e}}, \hspace*{2.5mm} e_\rf = e_\ri-u, \hspace*{2.5mm}
  \pi_\ri = (e_\ri^2 - \lambda^{-2})^{1/2}, \hspace*{2.5mm} \pi_\rf = (e_\rf^2 -
  \lambda^{-2})^{1/2}, \hspace*{2.5mm} \beta_\ri = 2\ln
  \frac{e_\ri+\pi_\ri}{\lambda^{-1}}, \hspace*{2.5mm} \beta_\rf = 2\ln
  \frac{e_\rf+\pi_\rf}{\lambda^{-1}},
\]
where $\alpha$ is the fine-structure constant, and i and f denote the
initial and final state of the electron, respectively.

Using these definitions, we can finally define the thermally averaged Gaunt
factor as
\begin{equation}
\langle g_{\rm ff}(\gamma^2,u,Z) \rangle = \frac{3\sqrt{6}}{32\sqrt{\upi}} \, \lambda^{7/2} \,
\frac{{\rm e}^u \, J^-(\lambda,\nu,u,Z)}{G^-_0(\lambda,\nu)} \hspace*{3mm} {\rm with}
\hspace*{3mm} \gamma^2 = \frac{Z^2 \, {\rm Ry}}{k T_{\rm e}} =
\frac{Z^2 \, {\rm Ry}}{m_{\rm e}c^2 \lambda} = \frac{(\alpha Z)^2}{2\lambda}
\approx 2.66257\times10^{-5} \frac{Z^2}{\lambda}.
\label{gff:rel}
\end{equation}
where Ry is the infinite-mass Rydberg unit of energy given by
\[ 1~{\rm Ry} = \alpha^2 m_{\rm e} c^2 / 2 \approx 2.17987\times10^{-18}~\rm{J}. \]

The numerical implementation of equations~(\ref{norm}) and (\ref{cs:int}) uses
similar techniques to what is described in paper~I. We used the same arbitrary
precision math libraries discussed in that paper, and all calculations were
done with the size of the mantissa fixed to 256 bits. This value was chosen
because 128 bits were found to be marginally insufficient near $\gamma^2 =
10^{10}$ and $u = 10^{-16}$. The integrals were evaluated using the adaptive
stepsize algorithm described in Sect.~3 of paper~I, which includes an estimate
for the error in the result. We assured that these estimates were correct by
comparing integrals computed with different values for the tolerance. This way
we assured that the relative numerical error of the result is always better
than $3.6\times10^{-5}$ in all tables, but typically the relative error will
be around $10^{-6}$. The tables were calculated using the same range of
parameters as in paper~I. This is ${}^{10}\log\gamma^2 = -6(0.2)10$ and
${}^{10}\log u = -16(0.2)13$. The notation $-6(0.2)10$ indicates that the
Gaunt factor was tabulated for all values of ${}^{10}\log\gamma^2$ ranging
from $-6$ to $10$ in increments of $0.2$ dex, and similarly for ${}^{10}\log
u$. Since the relativistic effects break the degeneracy in atomic number $Z$,
we calculated separate tables for all atomic numbers between $Z=1$ and $Z=36$.
These tables of pure BHE results are shown in Table~\ref{av} and are also
available in electronic form online. However, they will generally not be
directly usable, as is discussed in Sect.~\ref{merging:data}.

We compared our calculations to the data presented in Tables 1 through 4 of
N98 and found them to be in good agreement. The largest discrepancy was less
than 0.42\% for $\log\gamma^2 = -1.5$, $\log u = 0$, and $Z = 1$ where N98
found a Gaunt factor of $1.054$ and we found $1.05838$. The median discrepancy
is 0.111\% for Tables 1 and 2 of N98 and 0.260\% for Tables 3 and 4. Hence the
deviations are generally in good agreement with the relative error of 0.2\%
(for $Z \leq 8$) and 0.4\% (for heavier elements) claimed by N98.

\begin{table*}
\caption{$\langle g_{\rm ff}(\gamma^2,u,Z=1)\rangle$. This table shows an
  excerpt of the relativistic thermally averaged Gaunt factors that we
  calculated for $Z=1$. The full electronic version of this table, as well as
  tables for other values of $Z$, are available online. Over the parameter
  range shown in this table, the results are identical for the pure BHE and
  merged calculations. So only one table will be shown here, but both pure BHE
  and merged data sets are available online. Entries 1.50359115$+2$ mean
  $1.50359115 \times 10^{+2}$.\label{av}}
\begin{tabular}{rrrrrrr}
\hline
   & \multicolumn{6}{c}{${}^{10}\log \gamma^2$} \\
${}^{10}\log u$ & $-6.00$ & $-5.80$ &   $-5.60$ &    $-5.40$ &     $-5.20$ &     $-5.00$ \\
\hline
$-16.00$ & 1.50359115$+2$ & 1.18173936$+2$ & 9.29240645$+1$ & 7.31723286$+1$ & 5.78208822$+1$ & 4.60494480$+1$ \\
$-15.80$ & 1.48716575$+2$ & 1.16868580$+2$ & 9.18858822$+1$ & 7.23452413$+1$ & 5.71594368$+1$ & 4.55161151$+1$ \\
$-15.60$ & 1.47074038$+2$ & 1.15563191$+2$ & 9.08477193$+1$ & 7.15181690$+1$ & 5.64979904$+1$ & 4.49827813$+1$ \\
$-15.40$ & 1.45431516$+2$ & 1.14257820$+2$ & 8.98095399$+1$ & 7.06910870$+1$ & 5.58365349$+1$ & 4.44494486$+1$ \\
$-15.20$ & 1.43788950$+2$ & 1.12952445$+2$ & 8.87713637$+1$ & 6.98640156$+1$ & 5.51750929$+1$ & 4.39161154$+1$ \\
$-15.00$ & 1.42146437$+2$ & 1.11647069$+2$ & 8.77331844$+1$ & 6.90369343$+1$ & 5.45136379$+1$ & 4.33827811$+1$ \\
\hline
\end{tabular}
\end{table*}

\subsection{Merging relativistic and non-relativistic results}
\label{merging:data}

\begin{figure*}
\centerline{
\includegraphics[width=0.45\textwidth]{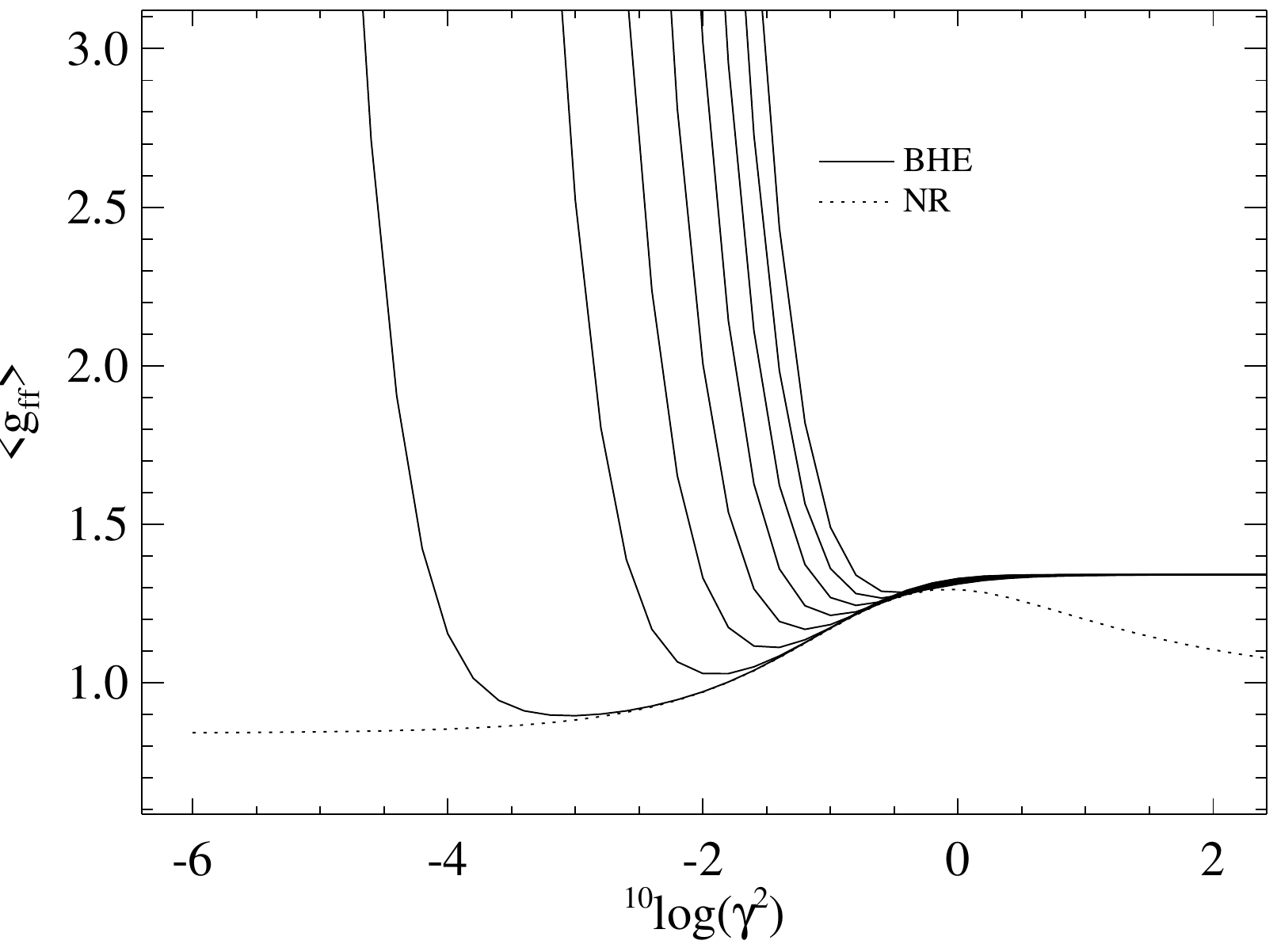}
\includegraphics[width=0.45\textwidth]{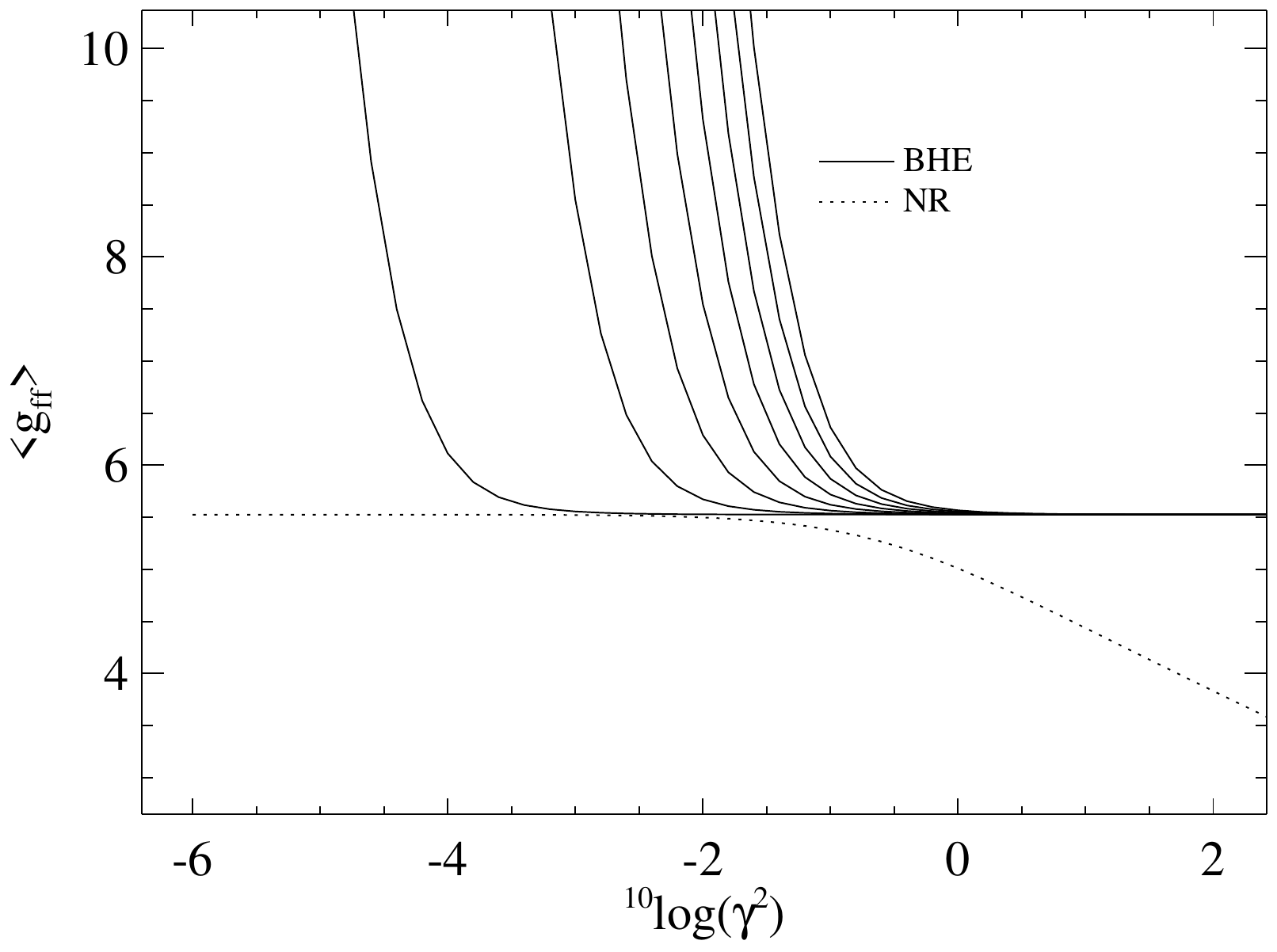}
}
\caption{These figures show the BHE data for $Z=1$ (left-most solid curve)
  through 36 (right-most solid curve) in increments of 5, for $u = 1$ (left
  panel) and $u = 10^{-4}$ (right panel). The exact non-relativistic results
  are indicated by the dotted line.
\label{elwert:cu}}
\end{figure*}

\begin{figure*}
\centerline{ \includegraphics[width=0.45\textwidth]{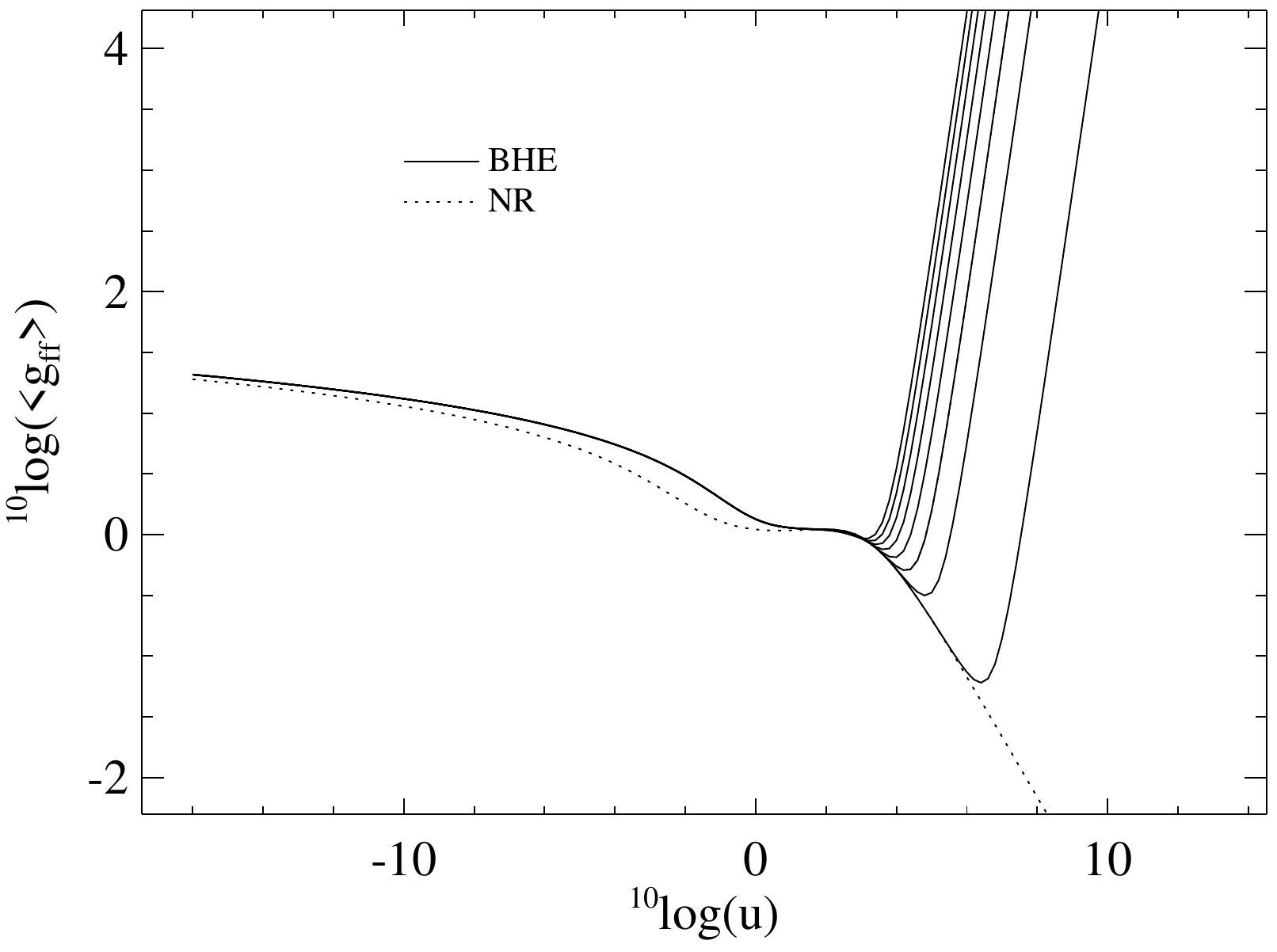}
  \includegraphics[width=0.45\textwidth]{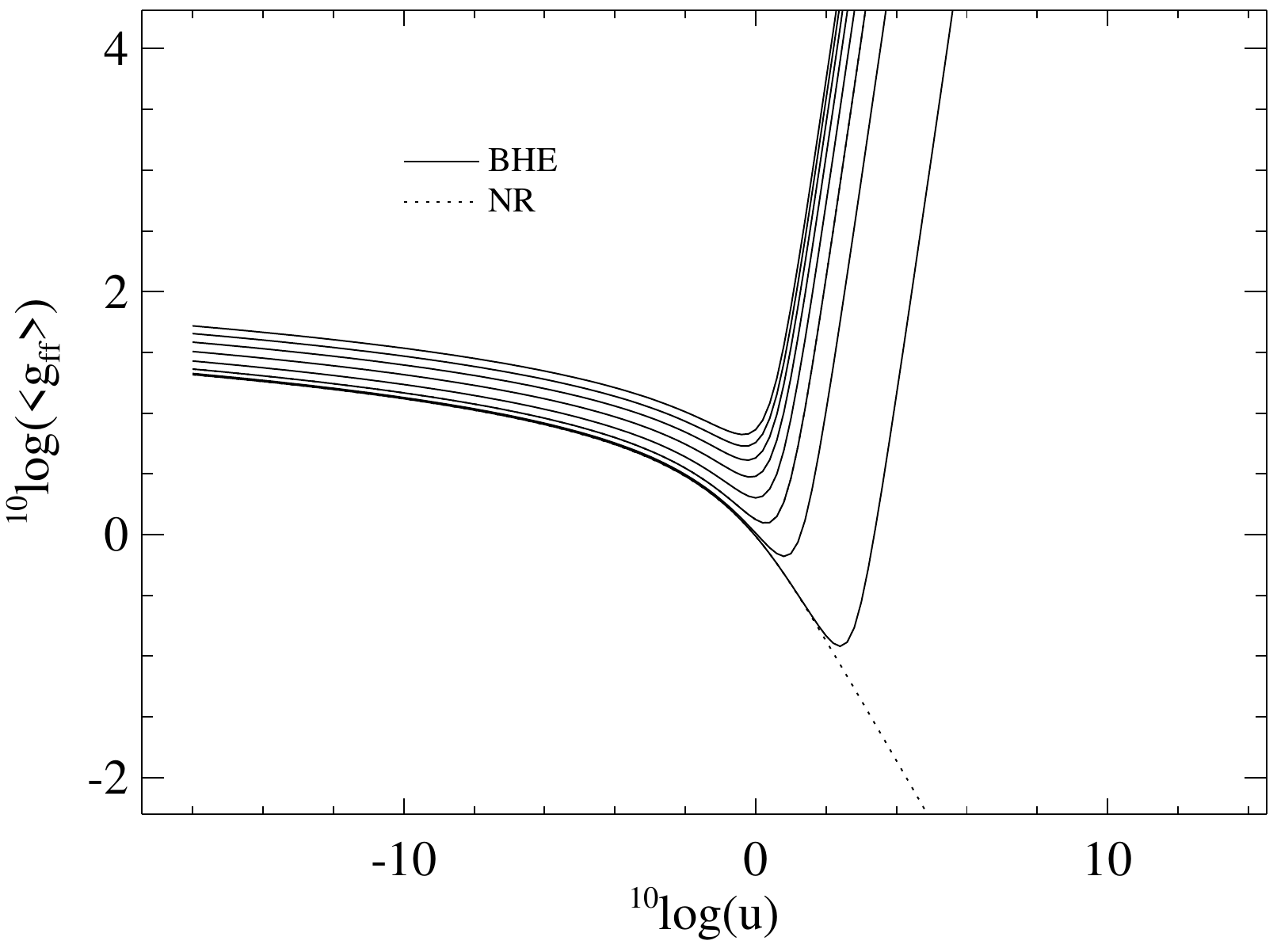} }
\caption{These figures show the BHE data for $Z=1$ (right-most solid curve)
  through 36 (left-most solid curve) in increments of 5, for $\gamma^2 = 100$
  (left panel) and $\gamma^2 = 10^{-2}$ (right panel). The exact
  non-relativistic results are indicated by the dotted line.
\label{elwert:cg}}
\end{figure*}

The results of the calculations discussed in the previous section can be seen
in Figs.~\ref{elwert:cu} and \ref{elwert:cg}. In Fig.~\ref{elwert:cu} we can
see that for low as well as high $\gamma^2$ values the NR and BHE calculations
disagree. For low $\gamma^2$ values (high temperatures) this is because the
assumptions in the non-relativistic calculation break down and the BHE results
should be used. For high $\gamma^2$ values (low temperatures) on the other
hand the Coulomb distortion of the wave functions becomes very large and the
Born approximation used by \citet{BH34} breaks down. In this regime the BHE
approximation cannot be used and the NR results should be adopted. In
Fig.~\ref{elwert:cg} we see that both for low and high $u$ values the NR and
BHE calculations disagree. For sufficiently large photon energies (larger than
roughly 100 -- 10\,000 Ry) this is again because the assumptions in the
non-relativistic calculation break down and the relativistic results should be
used. For lower photon energies the situation is more complex however. For low
temperatures (as shown in the left panel of Fig.~\ref{elwert:cg}) the
non-relativistic results will be more accurate and they should be used. For
high temperatures (as shown in the right panel of Fig.~\ref{elwert:cg}) on the
other hand the relativistic results should be used for all values of $u$.

\begin{figure*}
\centerline{
\includegraphics[width=0.45\textwidth]{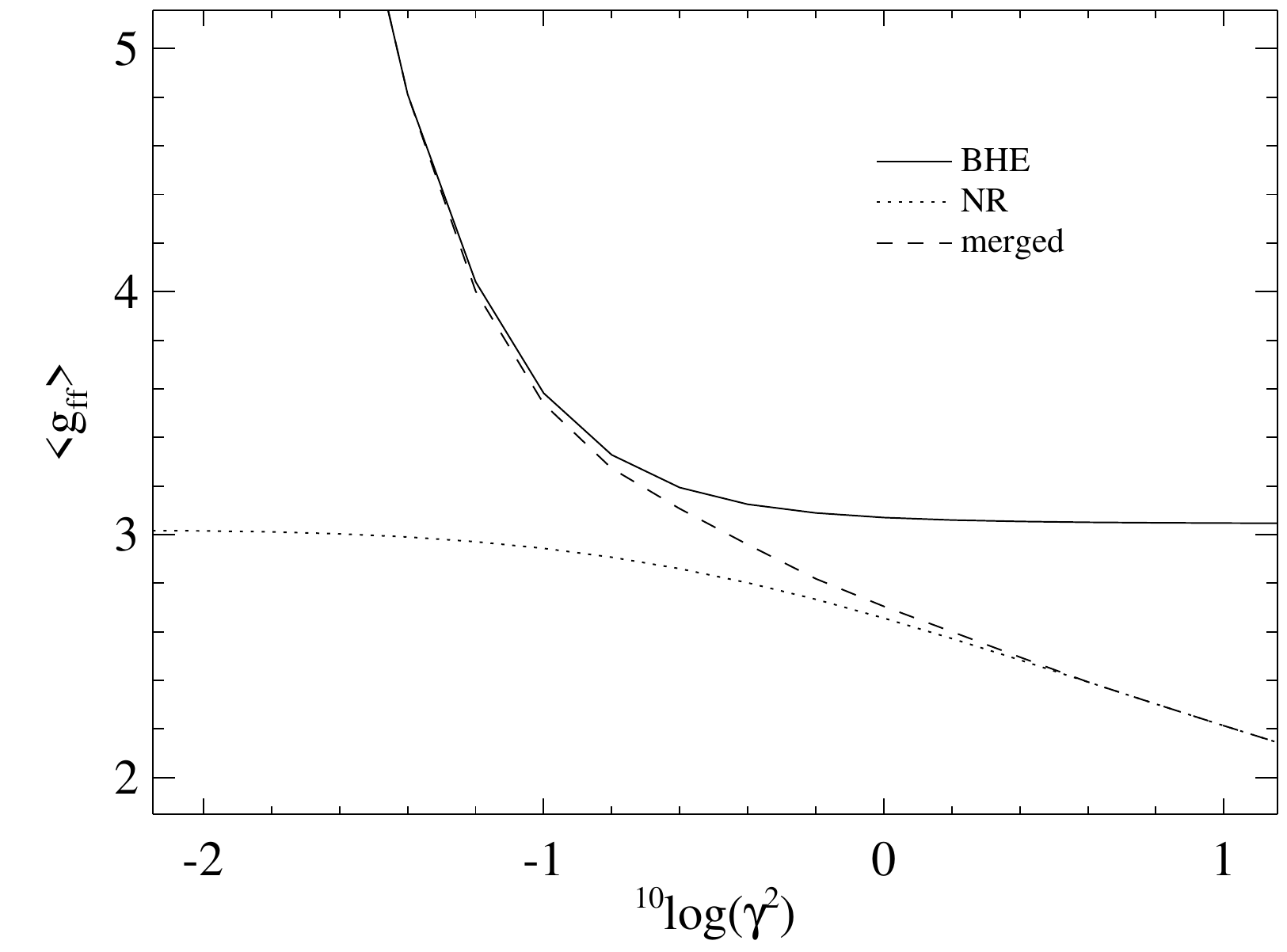}
\includegraphics[width=0.45\textwidth]{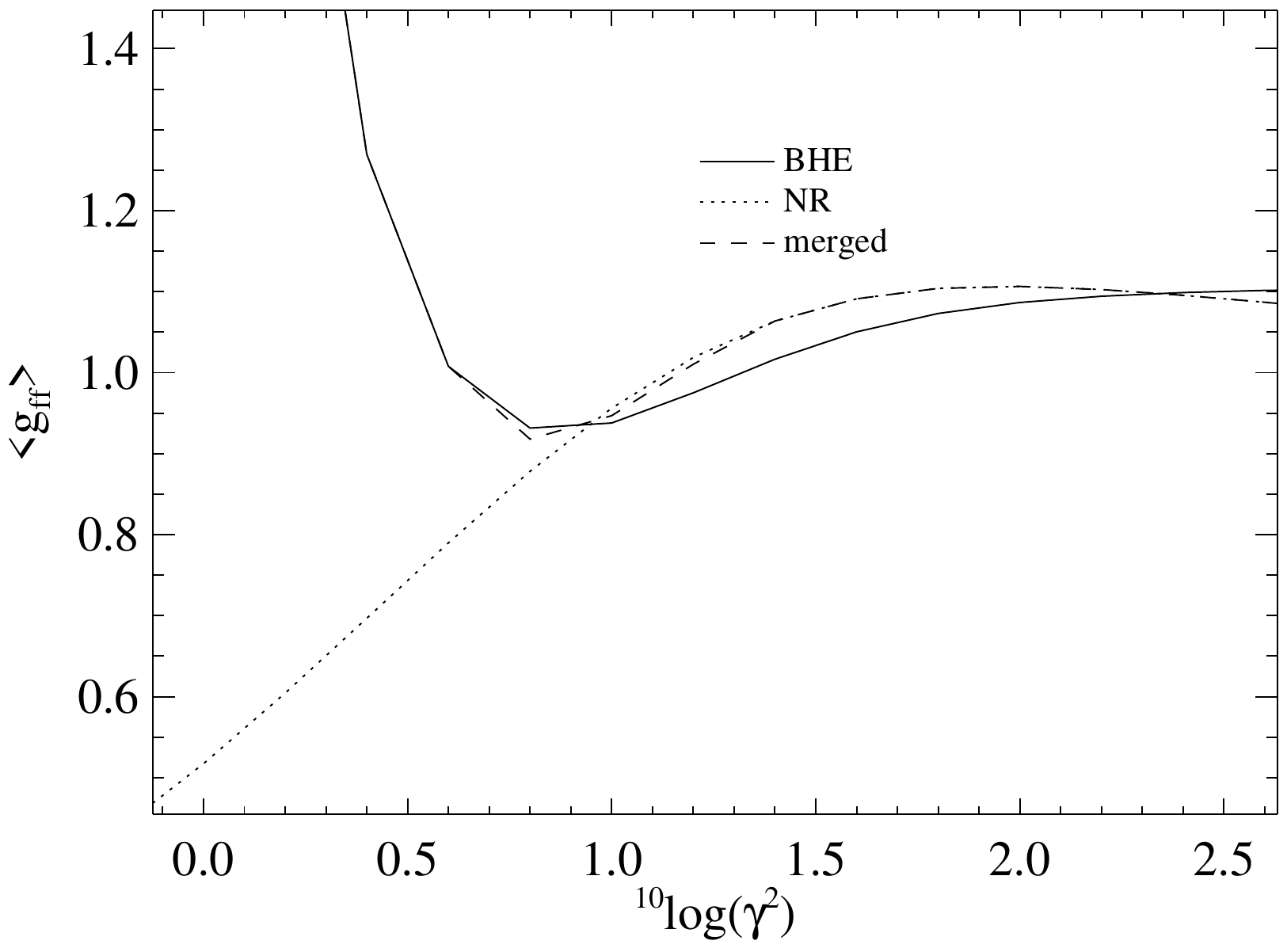}
}
\caption{These figures show the result of merging the BHE data for $Z=36$ with
  the exact non-relativistic data from paper~I. The left panel is for $u =
  0.01$ and the right panel for $u = 100$.
\label{merge}}
\end{figure*}

From this discussion it is clear that neither the NR nor the BHE results can
be used unchanged over the entire parameter range. We need to merge the
relativistic and non-relativistic results to obtain a data set that is
accurate for all values of $\gamma^2$, $u$, and $Z$. For this we use the
following algorithm. For every possible value of $u$ and $Z$ we compare both
data sets as a function of increasing values of $\gamma^2$. For every value of
$\gamma^2$ we compute the distance between the relativistic and
non-relativistic curve. When we view these results as a function of $\gamma^2$
one of the following three things can happen.
\begin{enumerate}
\item The curves never intersect, but the distance reaches a minimum value for
  a given $\gamma^2$. This case is shown in the left panel of
  Fig.~\ref{merge}. In this case we choose the change-over point as the
  $\gamma^2$ value where the distance is minimal. This happens for low $u$
  values (${}^{10}\log u < 0.6$ for $Z=1$).
\item The curves intersect in multiple places. This case is shown in the right
  panel of Fig.~\ref{merge}. In this case we choose the change-over point as
  the tabulated $\gamma^2$ value closest to the left-most intersect point.
  This happens for higher $u$ values ($0.6 \leq {}^{10}\log u \leq 12$ for
  $Z=1$).
\item Neither of these two things happen and the distance is monotonically
  decreasing without reaching either a minimum or intersect point. In this
  case no change-over point is chosen and the relativistic data are used for
  all values of $\gamma^2$. This happens for the highest $u$ values
  (${}^{10}\log u > 12$ for $Z=1$).
\end{enumerate}
If a change-over point was found, then the relativistic data will be used
below the change-over point and the non-relativistic data above. Immediately
around the change-over point a smooth transition from one curve to the other
will be created. The transition region will be between 3 and 9 tabulation
points wide, depending on how large the minimum distance is between the NR and
BHE curve. At the change-over point we will adopt the geometric mean of the NR
and BHE result: $(g_{\rm NR}^{\phantom{3}}g_{\rm BHE}^{\phantom{3}})^{1/2}$.
At the adjacent points we will use $(g_{\rm NR}^3g_{\rm
  BHE}^{\phantom{3}})^{1/4}$ and $(g_{\rm NR}^{\phantom{3}}g_{\rm
  BHE}^3)^{1/4}$, etc. The full details of the algorithm can be found in the
program {\sc merge.cc} which is included in the source tarball available on
the {\sc cloudy} web site.

\begin{figure*}
\centerline{
\includegraphics[width=0.45\textwidth]{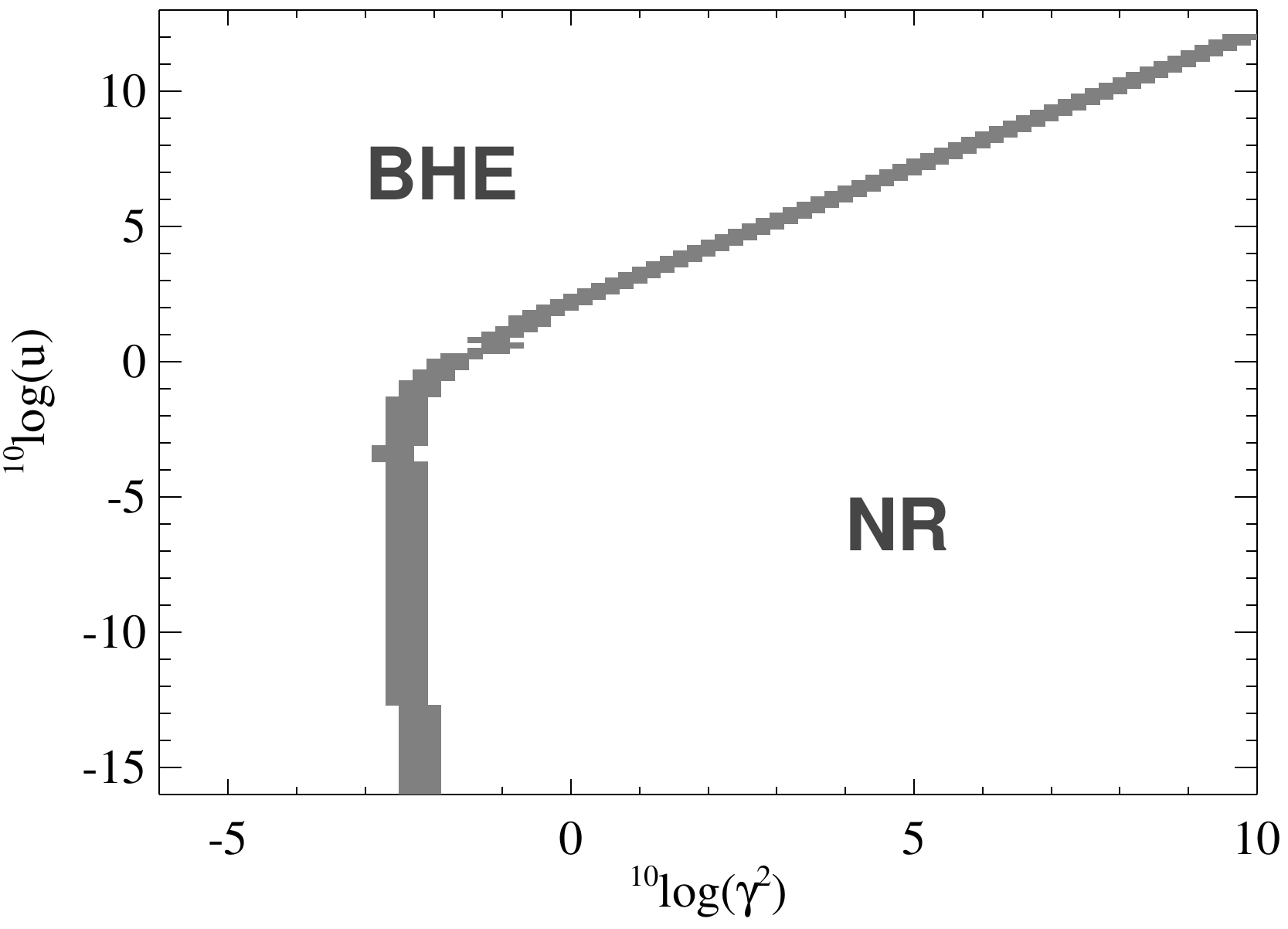}
\includegraphics[width=0.45\textwidth]{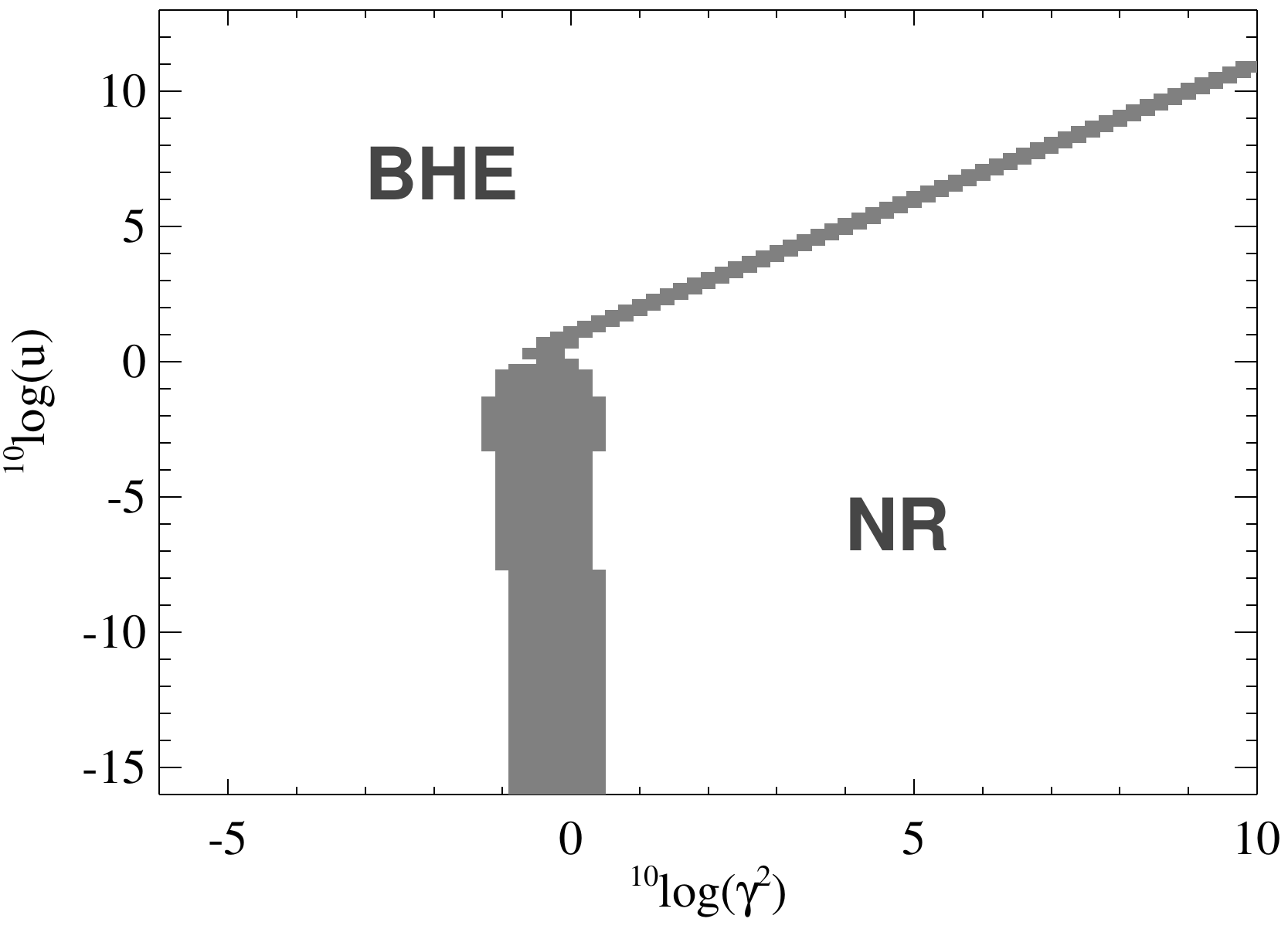}
}
\caption{These figures show the source of the Gaunt factor as a function of
  $\log\gamma^2$ and $\log u$ for $Z=1$ (left panel) and $Z=36$ (right panel).
  The shaded area is where the smooth transition between the BHE and NR data
  is created.
\label{merge:2d}}
\end{figure*}

\begin{figure*}
\centerline{ \includegraphics[width=0.45\textwidth]{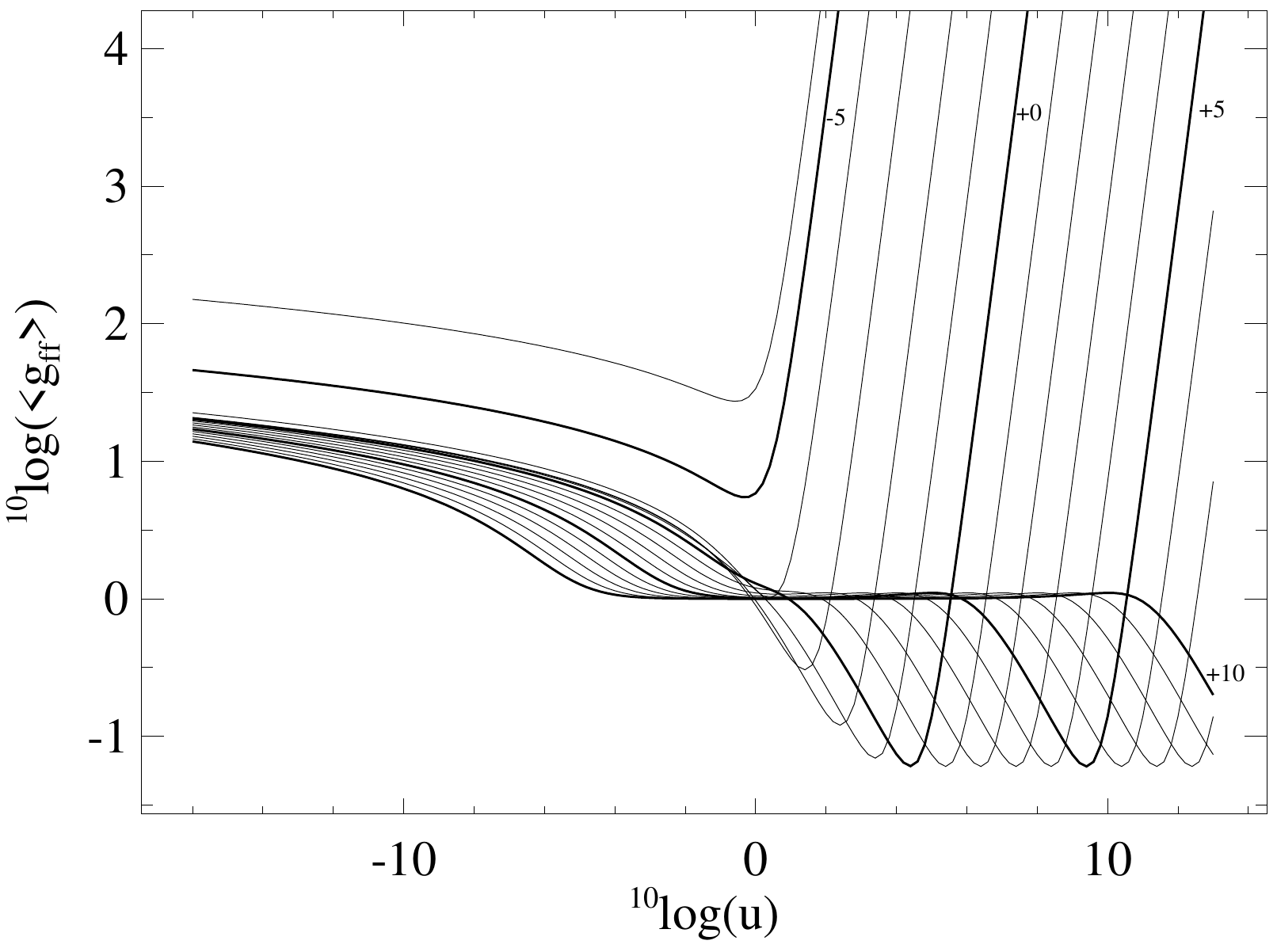}
  \includegraphics[width=0.45\textwidth]{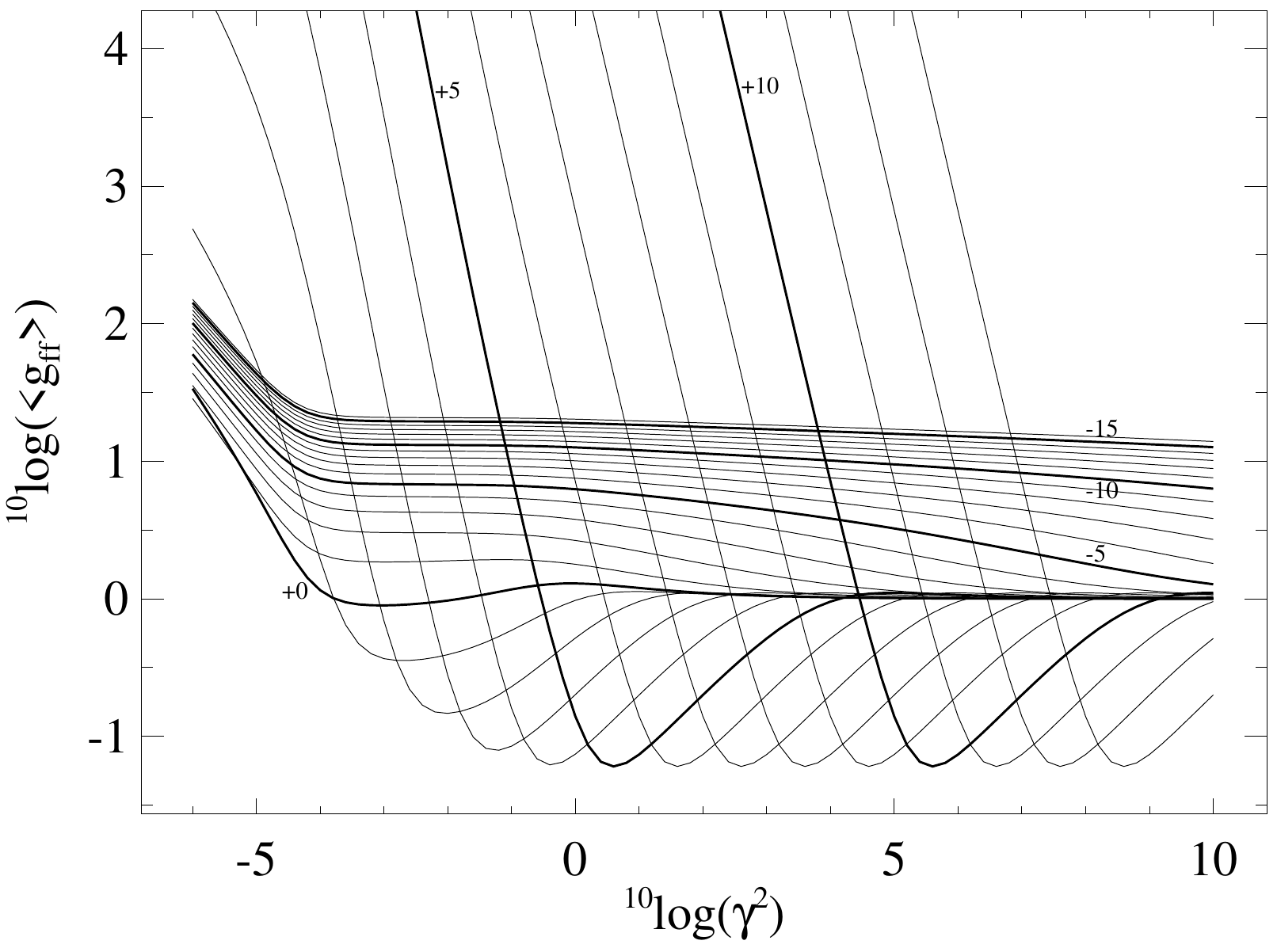} }
\caption{These figures show the merged free--free Gaunt data for Z=1 as a
  function of $u$ (left panel) and $\gamma^2$ (right panel). Thick curves are
  labelled with the values of ${}^{10}\log\gamma^2$ (left panel) and
  ${}^{10}\log u$ (right panel) in increments of 5 dex. The thin curves have a
  spacing of 1 dex.\label{merged:z01}}
\end{figure*}

\begin{figure*}
\centerline{
\includegraphics[width=0.45\textwidth]{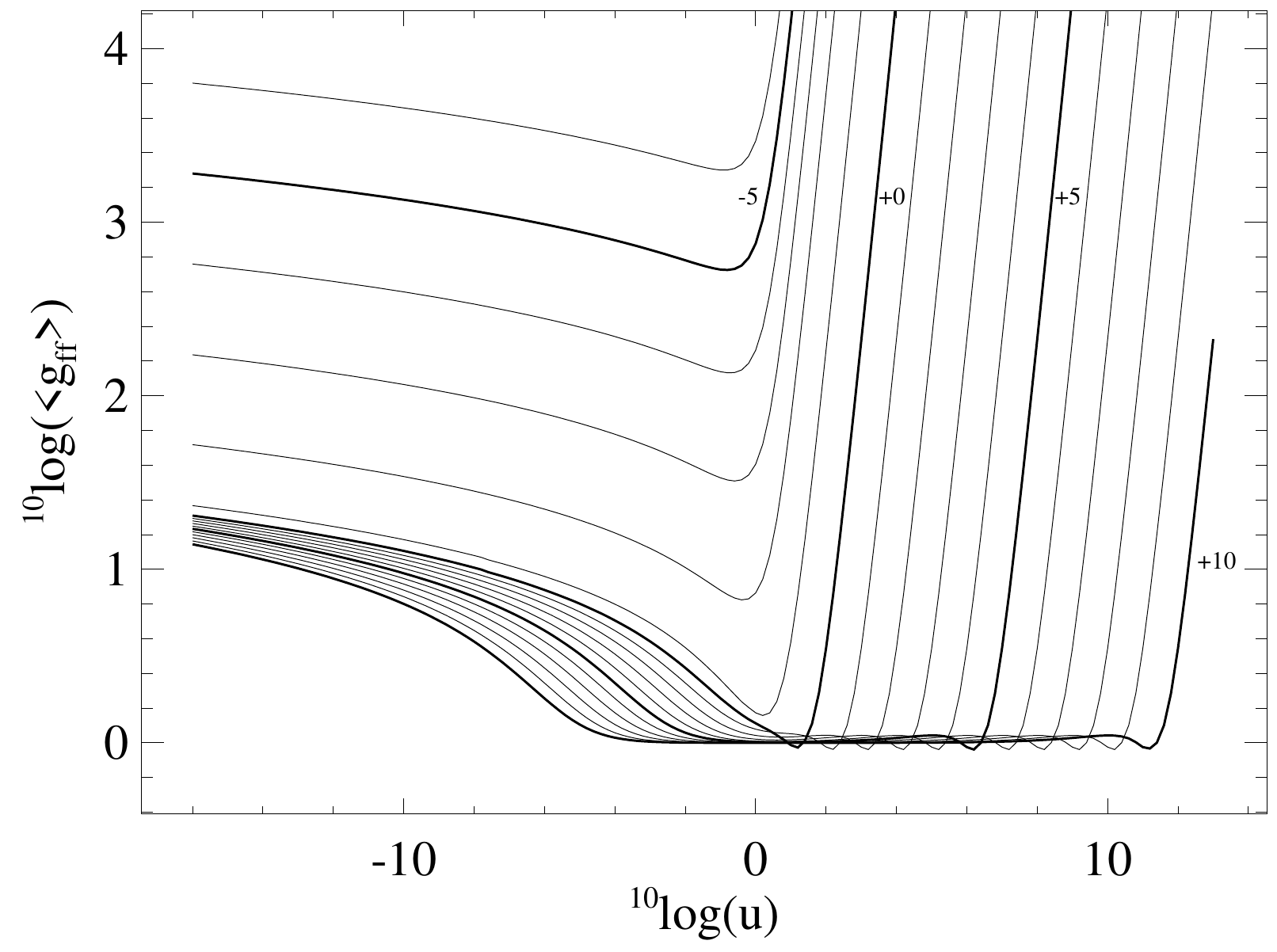}
\includegraphics[width=0.45\textwidth]{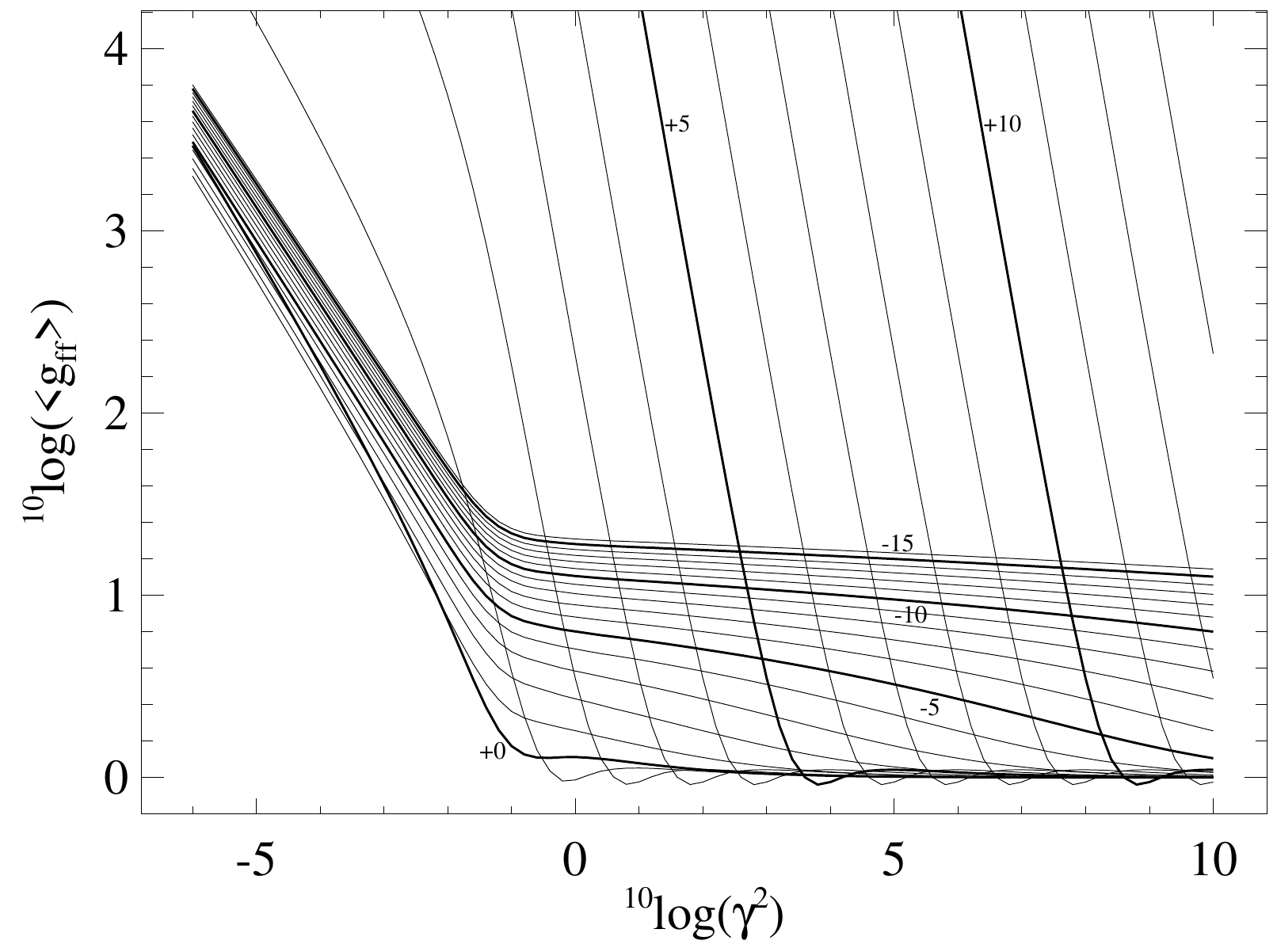}
}
\caption{Same as Fig.~\ref{merged:z01}, but for $Z = 36$.\label{merged:z36}}
\end{figure*}

The results of this algorithm can be seen in
Fig.~\ref{merge}. It should be noted that the case shown for $Z=36$ has the
worst match between the relativistic and non-relativistic results. For lower
$Z$ values the match will be better. The regions of the parameter space where
either the BHE or NR data are used are depicted graphically in
Fig.~\ref{merge:2d}. The resulting merged data sets are available online, and
are also shown in Table~\ref{av} and Figs.~\ref{merged:z01} and
\ref{merged:z36}. These tables should be used for plasma simulations.

\subsection{Spectral simulations with Cloudy}

\begin{figure*}
\centerline{
\includegraphics[width=0.65\textwidth]{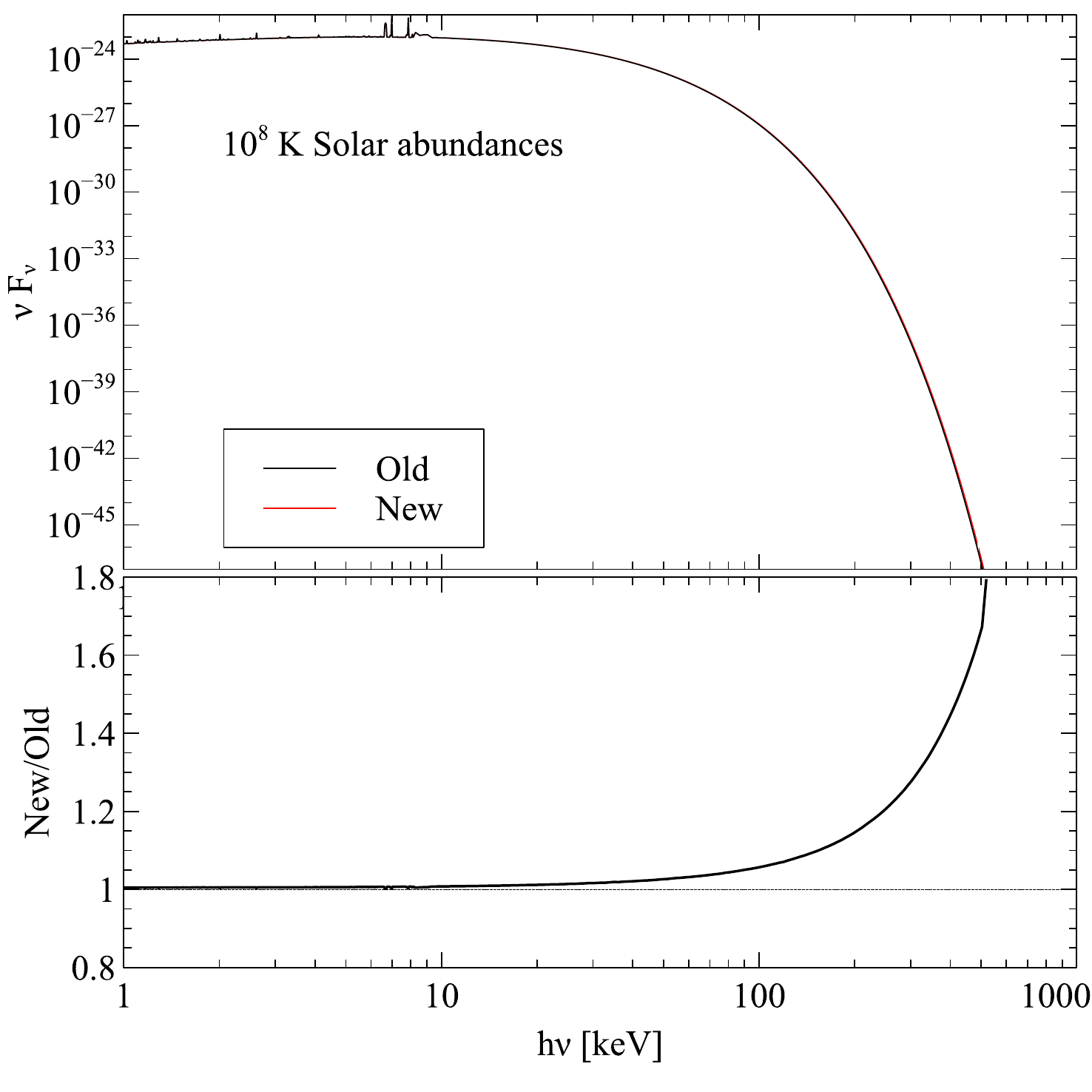}
}
\caption{The upper panel shows the spectrum of a 10$^8$~K gas with solar
  abundances in coronal equilibrium, computed both with the old and new
  treatments. The lower panel shows the ratio. Significant enhancements in the
  continuous emission are present at high energies.
\label{elwert:CoronalCompare.pdf}}
\end{figure*}

We have incorporated this improved theory into the development version of the
spectral simulation code {\sc cloudy}, which can simulate both photoionised
and collisionally ionised gas. The largest differences are expected at high
temperatures and photon energies. As an example of the effects of the improved
Gaunt factor we show the spectrum of a solar-abundance low-density gas with a
temperature of 100~MK in coronal equilibrium in
Fig.~\ref{elwert:CoronalCompare.pdf}. We compare the old Gaunt data in {\sc
  cloudy} version c13.03 (a combination of NR data with various
extrapolations) and the new data presented in this paper. The upper panel
shows the spectrum while the lower panel shows the ratio of new to old
treatments. Significant enhancements in the continuous emission occur at high
energies. The new data presented in this paper will be incorporated in the
next major release of {\sc cloudy}.

\section{The total free--free Gaunt factor}
\label{totalff}

Similar to paper~I, we will include a calculation of the total free--free
Gaunt factor which is integrated over frequency. Analytic fits to this
quantity were presented in \citet{It02} for $6.0 \leq T_{\rm e} \leq 8.5$ and
$1 \leq Z \leq 28$. The data presented here extend these results in $T_{\rm
  e}$ as well as $Z$. The formula for the frequency integrated Gaunt factor is
given by \citet{KL61}

\begin{equation}
\langle g_{\rm ff}(\gamma^2,Z)\rangle \; = \int_0^\infty {\rm e}^{-u} \langle g_{\rm ff}(\gamma^2,u,Z)\rangle {\rm d}u.
\label{frav:gff}
\end{equation}

\begin{figure*}
\centerline{
\includegraphics[width=0.45\textwidth]{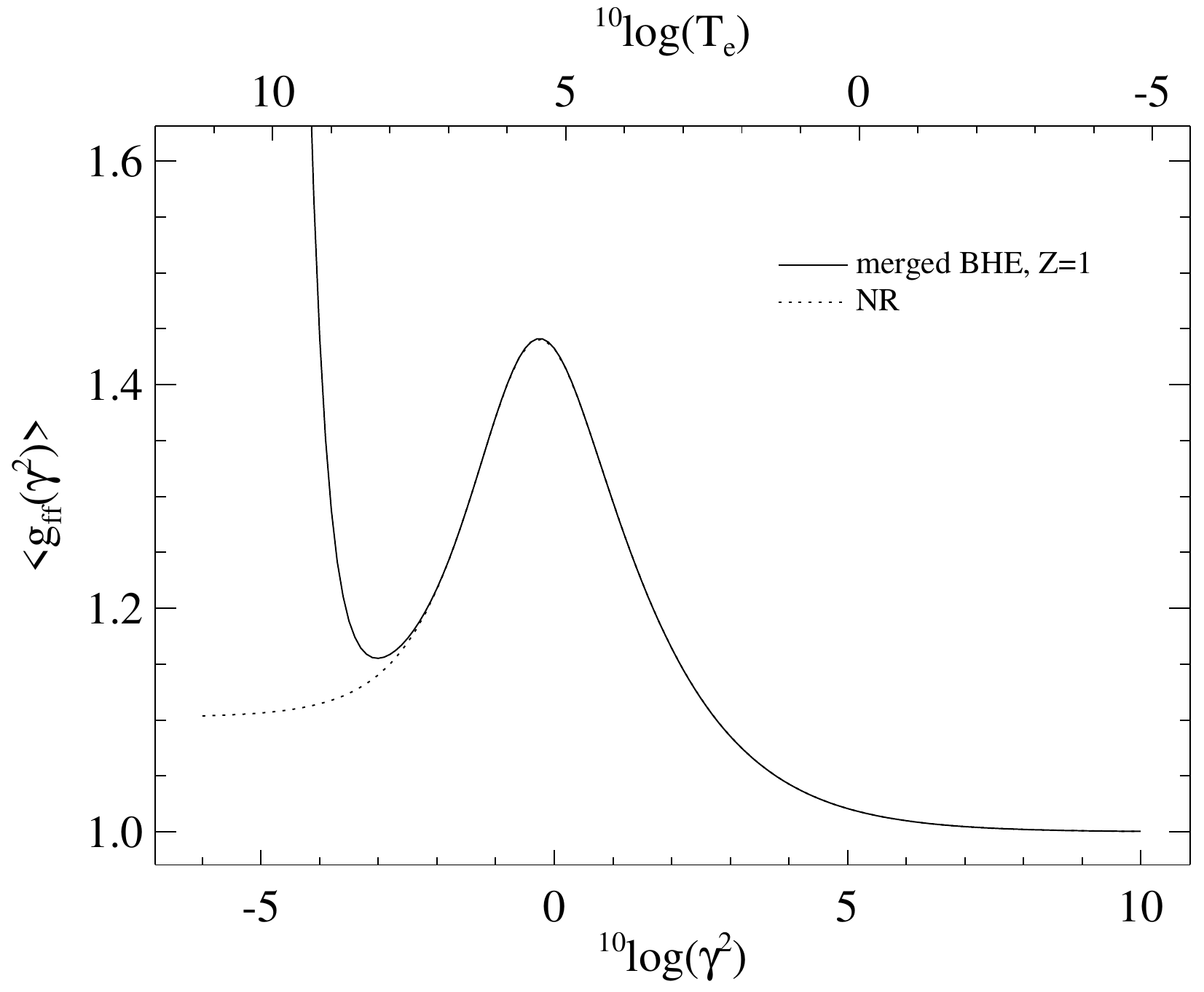}
\includegraphics[width=0.45\textwidth]{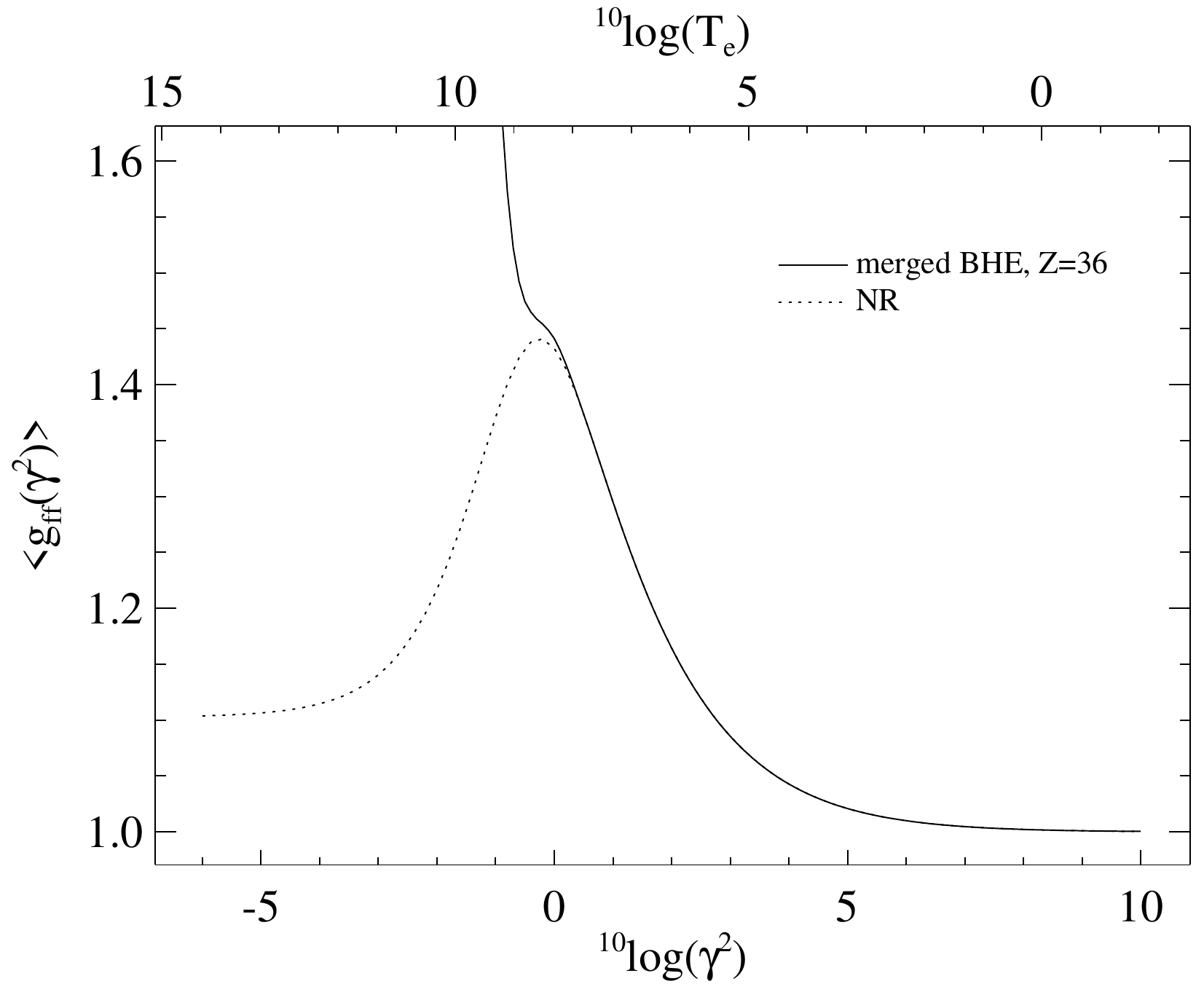}
}
\caption{The total free--free Gaunt factor as a function of $\gamma^2$ and
  $T_{\rm e}$. The left panel shows a comparison of the merged relativistic
  data for $Z=1$ (solid line) with the non-relativistic data taken from
  paper~I (dotted line). The right-hand panel shows the same, but for
  $Z=36$.\label{totgaunt}}
\end{figure*}

The resulting data are shown in Table~\ref{fravgff:tab}. This quantity is
useful for comparing the relativistic and non-relativistic Gaunt factors and
assess the magnitude of the relativistic effects as a function of temperature.
This comparison is shown in Fig.~\ref{totgaunt}. By inspecting these plots we
can see that for $Z=1$ the relativistic effects only become important above
$T_{\rm e} = 100$~MK. At 100~MK the relativistic effects increase the cooling
by slightly more than 0.75\% while the magnitude of the effect quickly rises
for higher temperatures. At 1~GK the relativistic effects raise the cooling by
more than 15\%, and at 10~GK by more than 317\%. For higher $Z$ species the
relativistic effects only become important at higher temperatures than for
lower $Z$ species. At $T_{\rm e} = 325$~MK the effect is still less than 1\%
for $Z=36$.

We calculated the total free--free Gaunt factor for all values of $Z$ between
1 and 36. These tables are available in electronic form on the {\sc cloudy}
website. We also provide simple {\sc fortran} and {\sc c} programs to
interpolate these tables.

\begin{table*}
\caption{This table shows an excerpt of the total free-free Gaunt factor as a
  function of $\gamma^2$ for $Z=1$. The full electronic version of this table,
  as well as tables for other values of $Z$ are available online. Entries
  3.92023$+1$ mean $3.92023 \times 10^{+1}$\label{fravgff:tab}}
\begin{tabular}{rr}
\hline
${}^{10}\log\gamma^2$ & $\langle g_{\rm ff}(\gamma^2,Z=1)\rangle$ \\
\hline
  $-6.00$ & 3.92023$+1$ \\
  $-5.90$ & 3.33120$+1$ \\
  $-5.80$ & 2.82268$+1$ \\
  $-5.70$ & 2.38583$+1$ \\
  $-5.60$ & 2.01184$+1$ \\
  $-5.50$ & 1.69078$+1$ \\
  $-5.40$ & 1.41710$+1$ \\
  $-5.30$ & 1.18333$+1$ \\
  $-5.20$ & 9.85344$+0$ \\
  $-5.10$ & 8.17565$+0$ \\
  $-5.00$ & 6.76918$+0$ \\
\hline
\end{tabular}
\end{table*}

\section{Summary}
\label{summary}

In this paper we presented calculations of the relativistic thermally averaged
Gaunt factor using the Bethe-Heitler-Elwert approximation. These data are not
valid for low temperatures and low photon energies because the Born
approximation used by \citet{BH34} breaks down in that regime. We have
therefore merged our data set with the non-relativistic data we presented in
paper~I. The BHE approximation is only valid for low values of $Z$, which is
why we have restricted our calculations to all values of $Z$ between 1 and 36.
A comparison of our calculations with the data presented by N98 showed that
they are in good agreement.

We also calculated the frequency integrated Gaunt factor for all values of $Z$
between 1 and 36. We compared these calculations with the non-relativistic
total Gaunt factor presented in paper~I. From this comparison we concluded
that relativistic effects only become important for electron temperatures in
excess of 100~MK and that relativistic effects are less pronounced for higher
$Z$ species at the same temperature.

All data presented in this paper are available in electronic form from MNRAS
as well as the {\sc cloudy} website at http://data.nublado.org/gauntff/. In
addition to these data tables, we also present simple interpolation routines
written in {\sc fortran} and {\sc c} on the {\sc cloudy} website. They use a
3rd-order Lagrange scheme to interpolate the logarithm of the thermally
averaged Gaunt data. This reaches a relative error better than
$3\times10^{-3}$ everywhere. The next release of {\sc cloudy} will contain a
vectorised version of the interpolation routine which is faster, while
maintaining the same precision. It is based on the Newton interpolation
technique. The program used to calculate all data is also available from the
{\sc cloudy} website.

\section*{Acknowledgements}

PvH acknowledges support from the Belgian Science Policy Office through the
ESA PRODEX program. GJF acknowledges support by NSF (1108928, 1109061, and
1412155), NASA (10-ATP10-0053, 10-ADAP10-0073, NNX12AH73G, and ATP13-0153),
and STScI (HST-AR-13245, GO-12560, HST-GO-12309, and GO-13310.002-A).

\bibliographystyle{mn2e}
\bibliography{gauntff}

\label{lastpage}

\end{document}